\documentclass[aps,prl,reprint,superscriptaddress]{revtex4-2}

\usepackage{graphicx}
\usepackage{dcolumn}
\usepackage{bm}
\usepackage{amsmath}
\usepackage{hyperref}
\hypersetup{colorlinks,breaklinks,linkcolor=blue,urlcolor=blue,anchorcolor=blue,citecolor=blue}

\begin{document}

\title{Grain-Growth Stagnation from Vacancy-Diffusion-Limited Disconnection Climb}

\author{Maik Punke}
 \affiliation{%
 Institute of Scientific Computing, TU Dresden, 01062 Dresden, Germany
}
 \author{Abel H. G. Milor}
\affiliation{%
 Institute of Scientific Computing, TU Dresden, 01062 Dresden, Germany
}
 \author{Marco Salvalaglio}%
\email{marco.salvalaglio@tu-dresden.de}
 \affiliation{%
 Institute of Scientific Computing, TU Dresden, 01062 Dresden, Germany
}
 \affiliation{Dresden Center for Computational Materials Science, TU Dresden, 01062 Dresden, Germany}

\date{\today}

\begin{abstract}
Grain growth in polycrystals typically stagnates at long times. We identify disconnection climb, limited by vacancy diffusion, as a fundamental microscopic mechanism underlying this behavior. Using a phase-field crystal framework extended to model vacancy diffusion, we resolve grain-boundary migration on diffusive time scales and show that disconnection climb rates correlate with the characteristic grain size at which growth arrests. These results link vacancy transport, disconnection dynamics, and microstructural evolution, establishing vacancy diffusion as a key governing factor.
\end{abstract}

\maketitle


Grain growth is a microstructural coarsening process in polycrystalline materials, during which grains of differing crystallographic orientation, separated by grain boundaries (GBs), increase in average size over time~\cite{burke1952recrystallization,von1952metal,mullins1956two,macpherson2007vonneumann}. This process is mainly driven by the tendency to reduce the total GB/interface energy and is importantly affected by the development of internal stresses associated with the shear displacement induced by GB motion~\cite{cahn2006coupling,thomas2017reconciling,qiu2025why,gautier2025quantifying}. Given the driving forces at play, this process should continue indefinitely, ultimately yielding a single crystal. However, real materials rarely reach this state, with grain growth slowing down and eventually stagnating at intermediate sizes.

Grain growth stagnation has been classically attributed to GB pinning effects at particles/pores~\cite{worner1992grain} and free-surface grooves ~\cite{mullins1958theeffect}, as well as to thickness effects in thin samples~\cite{burke1952recrystallization}.
However, it occurs even in high-purity, single-component bulk polycrystals~\cite{bolling1958grain,hu1974grain,holm2010grain}. Important explanations have been given in terms of the development of smooth, slow GBs~\cite{holm2010grain} and of disruptive atomic jumps at high temperature that affect GB structure~\cite{song2024disruptive}. 
This interplay between interfacial structure and emergent GB mobility directly points to the microscopic mechanisms governing GB migration. In single-component systems, their motion proceeds predominantly through the nucleation and propagation of steps carrying dislocation characters, i.e., disconnections~\cite{ashby1972boundary,hirth1973grain}. Disconnections glide over GBs in a diffusionless process, with mobility depending on GB structures~\cite{qiu2024grain}. In addition, similar to classical dislocations in bulk, they might also climb at high temperature and in the presence of vacancies~\cite{han2018grain}. Recent experimental atomic-scale observations indeed indicate that vacancy-mediated structural rearrangement of defects, and diffusion at defect cores, have a direct impact and can critically control GB migration~\cite{wei2021direct,wei2022direct,feng2023atomistic}.

Overall, grain-growth stagnation is a challenging phenomenon to control, whether the goal is to stabilize fine-grained microstructures through early stagnation (e.g., to achieve high strength) or to avoid it, and eventually fabricate a single crystal. Temperature is known to significantly influence grain-growth stagnation~\cite{holm2010grain}. Varying the temperature, however, affects activated processes, elastic constants, induces thermal strains, and alters the energetics and relative stabilities of both bulk and GB phases~\cite{han2016grain,Cantwell2020}. Consequently, temperature cannot be used to achieve controlled manipulation of grain-growth stagnation in otherwise equivalent microstructures.

\begin{figure*}[t]
\centering
\includegraphics[width=\linewidth]{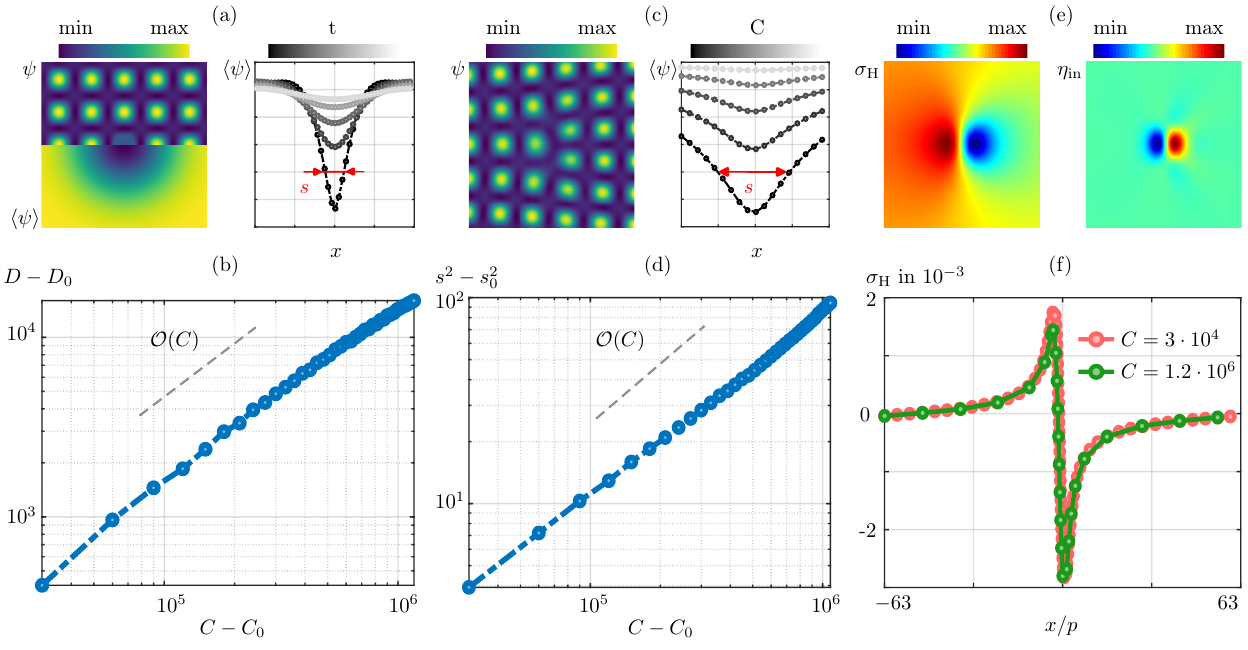}
\caption{\textit{Controlling vacancy diffusion}. (a) Crystal hosting a localized vacancy distribution illustrated by $\psi$ (left, upper half), $\langle \psi \rangle$ (left, lower half), and radial $\langle \psi \rangle$ over time (right). (b) Effective vacancy diffusivity $D=\partial s^2/(2\partial t)$ by varying $C$ with $C_0=3\cdot 10^4$, $D_0 = D(C_0)\approx 61$. 
(c) Crystal hosting a dislocation with Burgers vector aligned along $y$ axis (left) and $\langle \psi \rangle$ along the $x$ axis by varying $C$. (d) Effective diffusion length $s^2$ by varying $C$ with $s^2_0 = s^2(C_0) \approx 99$.
(e) Hydrostatic stress $\sigma_{\rm H}=\sum_i \sigma_{ii}/2$ (left) and stress incompatibility field $\eta_{\mathrm{in}}$ (right)~\cite{skogvoll2021dislocation} of the dislocation in panel (c). (f) $\sigma_{\rm H}(x,0)$ for two values of $C$ at the bound of the considered range in this work with lattice unit $p=2\pi$. Model parameters are: $\kappa = 1$, $T = 0.3$, $\psi_0 =-0.35$, $\rho = 1$ and $\Gamma = 10^4$.
}
	\label{fig:effVacDiff}
\end{figure*}

\begin{figure*}[t]
\centering
\includegraphics[width=\linewidth]{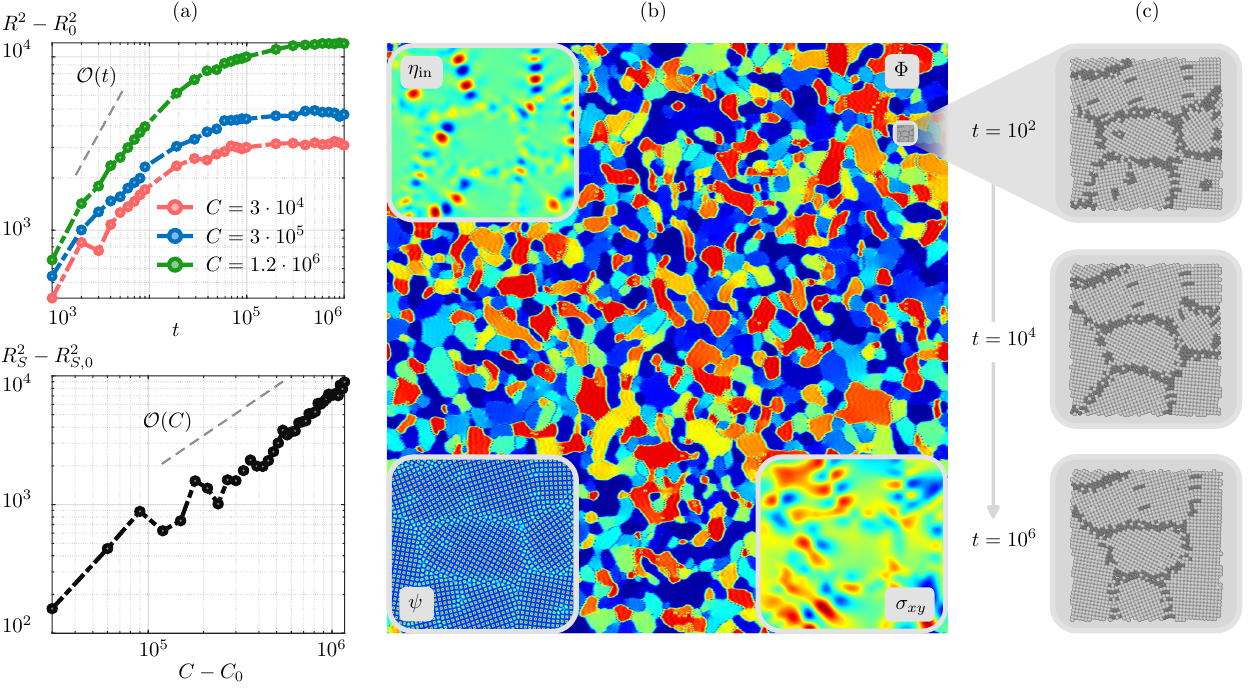}
\caption{\textit{The impact of vacancy diffusion on grain growth}. (a) Evolution of the mean-squared grain size ($R^2$, upper half) and mean-squared grain size of stagnating microstructure (lower half) for various values of $C$ with $R^2_0=R^2(t=0)\approx 1.6\cdot 10^{4}$, $R^2_{S,0}=R^2_S(C=C_0)\approx 1.9\cdot 10^{4}$ and $C_0=3\cdot 10^4$. 
(b) Microstructure under investigation visualized by the relative orientation of grains $\Phi$. Insets show the density field $\psi$ (bottom left), the shear stress $\sigma_{xy}$ (bottom right), and the stress incompatibility field $\eta_\mathrm{in}$ (top left). 
(c) Time-sequenced snapshots of a microstructure detail illustrating GB and dislocation migration obtained by reconstructed atom positions (using OVITO toolkit~\cite{stukowski2009visualization}) with model parameters as in Fig.~\ref{fig:effVacDiff}.}
	\label{fig:stagnation}
\end{figure*}
 
In this Letter, we outline the role of disconnection-climb–mediated GB migration over diffusive time scales in determining grain-growth stagnation. We isolate the effect of vacancy diffusion from other contributions, such as temperature, uncovering a potential new route to control this behavior. To access these physical aspects, we employ a phenomenological microscopic approach based on the phase field crystal (PFC) model~\cite{Elder2002}. Such a framework conveniently describes crystal structures over diffusive timescales, retaining lattice structures, elasticity, and defects~\cite{trautt2012coupled,Berry2014,skogvoll2022phase,qiu2024grain}, also reproducing grain growth stagnation~\cite{Bjerre2013,backofen2014capturing}, and is here extended to control vacancy diffusion. 

Consider a periodic order parameter $\psi\equiv\psi(\mathbf{x},t)$ mimicking a microscopic density field. We define the free energy
\begin{equation}
\label{eq:FullEnergy}
    \mathcal{F}[\psi, \langle\psi \rangle,\mathbf{v}] = \mathcal{F}_\mathcal{\psi}[\psi] + \int_\Omega \frac\psi2 | \mathbf{v} |^2 \mathrm{d\mathbf{x}}+ \int_\Omega \frac{C}{2} | \nabla \langle\psi \rangle |^2  \mathrm{d\mathbf{x}}.
\end{equation}
as an extension of the free energy entering the state-of-the-art hydrodynamic PFC formulation explicitly incorporating an elastic relaxation timescale~\cite{HeinonenPRL2016,skogvoll2022hydrodynamic,qiu2024grain}. The first term in Eq.~\eqref{eq:FullEnergy} consists of the minimal Swift-Hohenberg functional 
\begin{equation}
\label{eq:pfcEnergy}
\mathcal{F}_\mathcal{\psi}[\psi]= \int_{\Omega} \Big[\frac{\kappa}{2} \psi \mathcal{L}^2 \psi -\frac{T}{2} \psi^2+ \frac{\psi^4}{4} \Big] \,   \mathrm{d\mathbf{x}},
\end{equation}
with $\kappa$ a parameter controlling elastic constants as well as defect and interface energies, $T$ a phenomenological temperature~\cite{Elder2002}, and $\mathcal{L}$ a differential operator determining the periodicity of $\psi$ at equilibrium. For the sake of simplicity, we consider here 2D lattices with square crystalline symmetry (mimicking, e.g., bcc structures over \{001\} lattice planes) by setting $\mathcal{L}= \left(1+\nabla ^2 \right)\left(2+\nabla ^2 \right)$. 
The second term in Eq.~\eqref{eq:FullEnergy} introduces the kinetic energy associated with a macroscopic velocity field $\mathbf{v}$, varying over lengths significantly larger than the periodicity of $\psi$, that allows for fast wave-like propagation of atomic-density displacements  
~\cite{HeinonenPRL2016,skogvoll2022hydrodynamic}.
The third term in Eq.~\eqref{eq:FullEnergy} is newly introduced to effectively control vacancy diffusion via the coefficient $C$, as further motivated below. The operator $\langle \, \cdot \, \rangle$ represents a local spatial average (see its expression in the Supplemental Material (SM)~\cite{SupplementalMaterial}), with $\langle \psi \rangle \equiv \langle \psi \rangle(\mathbf{x},t)$ a slowly-varying density field that tracks variations of the average density on scales larger than the lattice spacing and is related to the vacancy concentration $\mathcal{V}\equiv \psi_0- \langle \psi \rangle$ with $\psi_0$ the density average over the whole system. 

Dynamics of $\psi$ and $\mathbf{v}$ are then obtained by combining linear response theory for relaxation toward equilibrium with the conservation of mass and momentum and read~\cite{skogvoll2022hydrodynamic}
\begin{equation}
\label{eq:pfcdynamics}
\begin{aligned}
        \partial_t \psi &= \nabla^2\dfrac{\delta \mathcal{F}_\psi[\psi]}{\delta \psi} - \mathbf{v}\cdot \nabla \psi- C \nabla^4 \langle \psi \rangle,  \\
    \rho \partial_t \mathbf{v} &= \Gamma \nabla^2 \mathbf{v} - \Big< \psi \nabla \dfrac{\delta  \mathcal{F}_\psi[\psi]}{\delta \psi}\Big>,
\end{aligned}
\end{equation}
with $\rho, \Gamma$ parameters controlling elastic relaxation with dissipation. Numerical solutions for these coupled equations are computed using a Fourier pseudo-spectral spatial discretization scheme. Details on the numerical method are reported in the SM~\cite{SupplementalMaterial}.

The last term in Eq.~\eqref{eq:FullEnergy} is motivated as follows. The classical PFC model~\cite{Elder2002,Emmerich2012} 
provides a coarse-grained, macroscopic description of vacancy diffusion: a crystalline region, described by a periodic $\psi$ whose amplitude is locally reduced (i.e., a localized distribution of $\mathcal{V}$), relaxes via diffusive dynamics toward a state with uniformly reduced amplitude~\cite{Elder2002}, such that $|\nabla \mathcal{V}| = |\nabla \langle \psi \rangle| \approx 0$ (Fig. \ref{fig:effVacDiff}a). This final state corresponds to vacancies effectively redistributed uniformly throughout the system. This process can then be modulated by penalizing or promoting $\nabla\langle \psi \rangle$, thereby controlling the timescale over which they evolve \footnote{The specific functional form follows as the simplest even term penalizing gradient, in analogy with classical phase field models}. At the same time, this term controls stationary deviations of $\langle \psi \rangle$ from $\psi_0$ at defects and interfaces (Fig. \ref{fig:effVacDiff}c). This behavior effectively mimics variation of concentration/segregation of vacancies at defects and GBs~\cite{estrin1982theory,McFadden2020}. Similar mathematical formulations have been used to control properties of defects and interfaces~\cite{guo2016interfacial,salvalaglio2017controlling,coelho2024generalizing}.

The effective vacancy diffusivity is found to scale linearly with $C$ above a certain value, see Fig.~\ref{fig:effVacDiff}b. This is estimated using the approach introduced in~\cite{Elder2004}: one density peak is removed to mimic a localized vacancy distribution. Its diffusion is monitored by fitting the evolving $\langle \psi \rangle$ profile to a Gaussian curve, from which the standard deviation $(s^2(t))$ is obtained. Since ordinary diffusion causes the width of a Gaussian profile to increase linearly in time at a rate determined by the diffusion constant, this time-dependence directly yields the effective vacancy diffusion constant $D=\partial s^2/(2\partial t)$, see Fig~\ref{fig:effVacDiff}a. We remark that the adopted procedure is valid for a given $T$ as, owing to its phenomenological character, diffusion of perturbation of the perfect crystal does not faithfully reproduce expected vacancy behavior with variation of this parameter~\cite{van2013vacancy}, which also affects elastic moduli and phase stability. However, note that the formulation considered here can be used to correct this behavior by properly defining $C(T)$.  The spreading of local average density profiles at defects is also found to scale linearly with $C$, as shown in Fig.~\ref{fig:effVacDiff}d.  $\langle \psi \rangle$ is computed at the dislocation, fitted to a Gaussian function, and its standard deviation $s^2$ is interpreted as the effective diffusion length over which vacancies spread into the bulk. 
Importantly, changing $C$ does not alter the long-range deformation induced by defects, see Figs.~\ref{fig:effVacDiff}e-\ref{fig:effVacDiff}f: the far-field stress governing the elastic interaction between defects remains unchanged with minor deviations at the core only.

Grain growth simulations are illustrated in Fig.~\ref{fig:stagnation}. They initially proceed resembling the expected linear scaling of the average grain size over time, $R^2\sim t$~\cite{macpherson2007vonneumann, qiu2025why} (Fig.~\ref{fig:stagnation}a, top). This quantity is estimated here using a newly introduced, high-fidelity method based on local grain orientations and persistent homology~\cite{edelsbrunner2022computational}, further detailed in the SM~\cite{SupplementalMaterial}.
At later stages, the dynamics slows down and eventually stops at a finite stagnation radius, consistent with experimental evidence as classically observed in other atomistic simulations~\cite{holm2010grain,Bjerre2013,backofen2014capturing} (Fig.~\ref{fig:stagnation}a, top). The size at which the process slows down and stagnation occurs increases linearly with this parameter for $C\gtrsim 3\cdot 10^4$ (Fig.~\ref{fig:stagnation}a, bottom), while it does not vary significantly below this value. 
Since $C$ exclusively controls the diffusive evolution of vacancies and their spreading at defects, as discussed above and illustrated in Fig.~\ref{fig:effVacDiff}, we obtain a direct correlation between vacancy diffusion and the onset of grain-growth stagnation as well as the stagnating average grain size. 

\begin{figure*}[t]
\centering
\includegraphics[width=\linewidth]{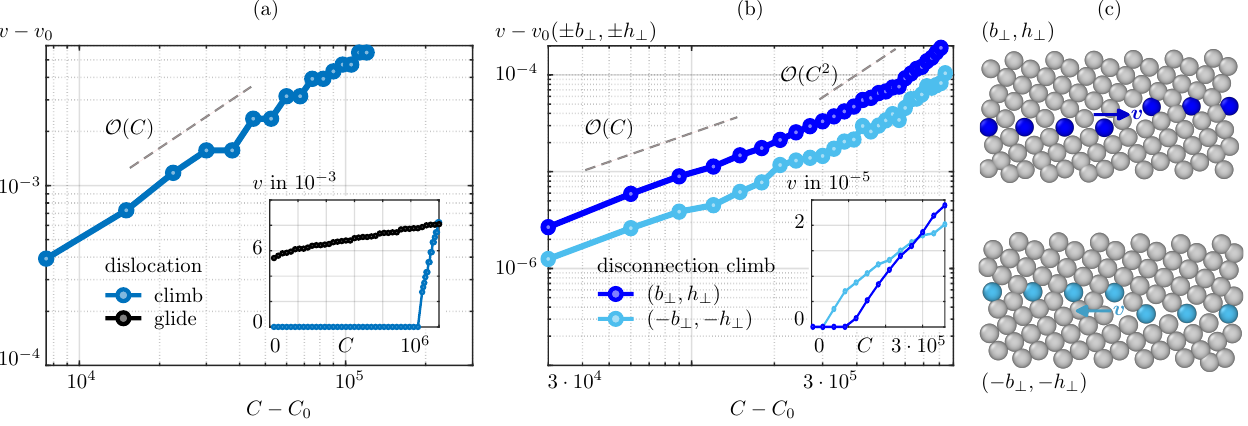}
\caption{\textit{The impact of vacancy diffusion on defect motion}. (a) Dislocation glide and climb velocity for various values of $C$ with $v_0=v(C_0)\approx 2.7\cdot 10^{-3}$ and $C_0=1.08\cdot 10^6$ (linear scale as inset).
(b) Disconnection climb velocity for various values of $C$ with $v_0(C=C_0)\approx 2.4\cdot 10^{-5}$ for $(b_\perp, -h_\perp)$, $v_0(C=C_0)\approx 2.0\cdot 10^{-5}$ for $(-b_\perp, -h_\perp)$ and $C_0=3.6\cdot 10^5$ (linear scale as inset)
(c) Close-up of the atomic structure in a GB region hosting disconnections with Burgers vector and step height $(b_\perp, h_\perp)$ (upper half) and $(-b_\perp, -h_\perp)$ (lower half). Reconstructed atom positions are obtained using OVITO toolkit~\cite{stukowski2009visualization} with model parameters as in Fig.~\ref{fig:effVacDiff}.}
\label{fig:defects}
\end{figure*}

To substantiate the interpretation of this effect as a consequence of controlled vacancy diffusion and its impact on GB migration, it remains to clarify and assess how variations in $C$ influence the underlying microscopic mechanisms. We recall that GB migration is mediated by flows of disconnections~\cite{han2018grain,qiu2025why,gautier2025quantifying}. These can be seen in the considered framework via the incompatible stress, which points to defects at GBs with dislocation character (Fig.~\ref{fig:stagnation}b). Moreover, lattice dislocations also nucleate, migrate, and impinge GBs (see Fig.~\ref{fig:stagnation}c). By modifying how vacancies are redistributed along and across GBs, $C$ is thus expected to regulate whether defect climb occurs and the overall speed of the process. 

We consider first the case of two edge dislocations annihilating through pure glide or pure climb (Fig.~\ref{fig:defects}a). The dislocation glide velocity exhibits only a weak dependence on $C$, consistent with the fact that vacancy diffusion is not a rate-limiting mechanism for glide and affects the process only indirectly, e.g., through minor modifications of the density field at the dislocation core. By contrast, $C$ has a pronounced effect on dislocation climb. For sufficiently low vacancy mobility ($C < 1.08\times10^{6}$), no measurable climb is observed, indicating that a minimum vacancy diffusion is required to sustain continuous climb motion. Above this threshold, the climb velocity increases approximately linearly with $C$. This result shows consistency with the known behavior for dislocations~\cite{anderson2017} and demonstrates a substantial improvement in PFC modeling, in particular concerning the control over dislocation climbing rates.

Next, we specifically analyze disconnection motion in a representative setting, namely along a symmetric-tilt $\Sigma 5$ GB (using classical Coincidence Site Lattice (CSL)~\cite{SuttonBalluffi} nomenclature). We initialize a disconnection dipole that can annihilate via climb and measure climb velocities of the two disconnections forming the dipole as a function of $C$ (Fig.~\ref{fig:defects}b). Analogously to bulk dislocation climb, these velocities become nonzero above a threshold value, comparable to the values at which the stagnating grain size begins to scale with $C$. Above such a threshold, velocities increase nearly linearly with this parameter. They then display an approximately linear-to-quadratic dependence over the explored range, which is ascribed to the superposition of the effect illustrated in Fig.~\ref{fig:effVacDiff}c for both the GB and the disconnection. Interestingly, the two velocities are different and the corresponding onsets of motion by climb occur at different values of $C$. The two disconnections possess identical Burgers-vector magnitudes and step heights, and are therefore subject to driving forces of equal magnitude and opposite sign. Nevertheless, they move in opposite directions over the GB (see Fig.~\ref{fig:defects}c).
Their different mobilities can be attributed to the fact that the $\Sigma 5$ GB is not mirror-symmetric with respect to the direction perpendicular to the GB. This result further shows that CSL-based bicrystallography is insufficient to capture the kinetics of GB motion, complementing recent evidence~\cite{zhang2020grain,qiu2024grain,xu2026scripta}. Notably, disconnections require substantially lower vacancy mobility to activate climb than lattice dislocations. This is consistent with their smaller Burgers vector compared to that of a lattice dislocation, implying weaker lattice distortion and reduced vacancy accumulation, combined with the intrinsic excess volume available at GBs that confine disconnections. 

In summary, we reported on grain growth at diffusive timescales and showed that vacancy diffusion governs both the onset of grain-growth stagnation and the resulting average grain size. Vacancy diffusion is explicitly shown to enable disconnection climb and to control its rate. When diffusion is limited, climb is suppressed, and GB motion proceeds primarily via disconnection glide; disconnections that require climb then act as effective pinning sites, hindering further grain coarsening. As vacancy diffusion increases, climb becomes progressively activated along GBs, enabling enhanced coarsening and ultimately larger grain sizes. This gradual transition can be attributed to the distribution of disconnections and disconnection modes \cite{han2018grain} across the GB network, for which climb becomes active under different conditions, as well as to the emergence of an effective diffusion length. This study also introduces a substantial improvement in scale-bridging models of microstructure evolution: compared to existing formulations, the considered PFC model enables control over defect climb rates and, for the first time, explicitly addresses disconnections. Overall, these results reveal a fundamental mechanism governing grain growth and its stagnation, suggesting new strategies for controlling these phenomena and providing a basis for extension to more complex systems, such as polycrystalline thin films and alloys, and in addition to known mechanisms leading to stagnation \cite{worner1992grain,mullins1958theeffect,holm2010grain}.

\textit{Acknowledgements} -- We thank Alex Mamaev for fruitful discussions. We acknowledge support from the Deutsche Forschungsgemeinschaft (DFG, German Research Foundation), project numbers 447241406, 493401063 (Research Training Group (GRK) 2868 D3), 570666382. We also gratefully acknowledge the computing resources on the high-performance computer at the NHR Center of TU Dresden. 

\textit{Data availability} -- The data and codes that support the findings of this article will be made openly available upon acceptance for publication.

\makeatletter
\providecommand \@ifxundefined [1]{%
 \@ifx{#1\undefined}
}%
\providecommand \@ifnum [1]{%
 \ifnum #1\expandafter \@firstoftwo
 \else \expandafter \@secondoftwo
 \fi
}%
\providecommand \@ifx [1]{%
 \ifx #1\expandafter \@firstoftwo
 \else \expandafter \@secondoftwo
 \fi
}%
\providecommand \natexlab [1]{#1}%
\providecommand \enquote  [1]{``#1''}%
\providecommand \bibnamefont  [1]{#1}%
\providecommand \bibfnamefont [1]{#1}%
\providecommand \citenamefont [1]{#1}%
\providecommand \href@noop [0]{\@secondoftwo}%
\providecommand \href [0]{\begingroup \@sanitize@url \@href}%
\providecommand \@href[1]{\@@startlink{#1}\@@href}%
\providecommand \@@href[1]{\endgroup#1\@@endlink}%
\providecommand \@sanitize@url [0]{\catcode `\\12\catcode `\$12\catcode
  `\&12\catcode `\#12\catcode `\^12\catcode `\_12\catcode `\%12\relax}%
\providecommand \@@startlink[1]{}%
\providecommand \@@endlink[0]{}%
\providecommand \url  [0]{\begingroup\@sanitize@url \@url }%
\providecommand \@url [1]{\endgroup\@href {#1}{\urlprefix }}%
\providecommand \urlprefix  [0]{URL }%
\providecommand \Eprint [0]{\href }%
\providecommand \doibase [0]{https://doi.org/}%
\providecommand \selectlanguage [0]{\@gobble}%
\providecommand \bibinfo  [0]{\@secondoftwo}%
\providecommand \bibfield  [0]{\@secondoftwo}%
\providecommand \translation [1]{[#1]}%
\providecommand \BibitemOpen [0]{}%
\providecommand \bibitemStop [0]{}%
\providecommand \bibitemNoStop [0]{.\EOS\space}%
\providecommand \EOS [0]{\spacefactor3000\relax}%
\providecommand \BibitemShut  [1]{\csname bibitem#1\endcsname}%
\let\auto@bib@innerbib\@empty
\bibliography{bibliography}

\begin{thebibliography}{48}%
\makeatletter
\providecommand \@ifxundefined [1]{%
 \@ifx{#1\undefined}
}%
\providecommand \@ifnum [1]{%
 \ifnum #1\expandafter \@firstoftwo
 \else \expandafter \@secondoftwo
 \fi
}%
\providecommand \@ifx [1]{%
 \ifx #1\expandafter \@firstoftwo
 \else \expandafter \@secondoftwo
 \fi
}%
\providecommand \natexlab [1]{#1}%
\providecommand \enquote  [1]{``#1''}%
\providecommand \bibnamefont  [1]{#1}%
\providecommand \bibfnamefont [1]{#1}%
\providecommand \citenamefont [1]{#1}%
\providecommand \href@noop [0]{\@secondoftwo}%
\providecommand \href [0]{\begingroup \@sanitize@url \@href}%
\providecommand \@href[1]{\@@startlink{#1}\@@href}%
\providecommand \@@href[1]{\endgroup#1\@@endlink}%
\providecommand \@sanitize@url [0]{\catcode `\\12\catcode `\$12\catcode
  `\&12\catcode `\#12\catcode `\^12\catcode `\_12\catcode `\%12\relax}%
\providecommand \@@startlink[1]{}%
\providecommand \@@endlink[0]{}%
\providecommand \url  [0]{\begingroup\@sanitize@url \@url }%
\providecommand \@url [1]{\endgroup\@href {#1}{\urlprefix }}%
\providecommand \urlprefix  [0]{URL }%
\providecommand \Eprint [0]{\href }%
\providecommand \doibase [0]{https://doi.org/}%
\providecommand \selectlanguage [0]{\@gobble}%
\providecommand \bibinfo  [0]{\@secondoftwo}%
\providecommand \bibfield  [0]{\@secondoftwo}%
\providecommand \translation [1]{[#1]}%
\providecommand \BibitemOpen [0]{}%
\providecommand \bibitemStop [0]{}%
\providecommand \bibitemNoStop [0]{.\EOS\space}%
\providecommand \EOS [0]{\spacefactor3000\relax}%
\providecommand \BibitemShut  [1]{\csname bibitem#1\endcsname}%
\let\auto@bib@innerbib\@empty
\bibitem [{\citenamefont {Burke}\ and\ \citenamefont
  {Turnbull}(1952)}]{burke1952recrystallization}%
  \BibitemOpen
  \bibfield  {author} {\bibinfo {author} {\bibfnamefont {J.}~\bibnamefont
  {Burke}}\ and\ \bibinfo {author} {\bibfnamefont {D.}~\bibnamefont
  {Turnbull}},\ }\bibfield  {title} {\bibinfo {title} {Recrystallization and
  grain growth},\ }\href
  {https://doi.org/https://doi.org/10.1016/0502-8205(52)90009-9} {\bibfield
  {journal} {\bibinfo  {journal} {Prog. Metal Phys.}\ }\textbf {\bibinfo
  {volume} {3}},\ \bibinfo {pages} {220} (\bibinfo {year} {1952})}\BibitemShut
  {NoStop}%
\bibitem [{\citenamefont {Von~Neumann}(1952)}]{von1952metal}%
  \BibitemOpen
  \bibfield  {author} {\bibinfo {author} {\bibfnamefont {J.}~\bibnamefont
  {Von~Neumann}},\ }\bibfield  {title} {\bibinfo {title} {Metal interfaces},\
  }\href@noop {} {\bibfield  {journal} {\bibinfo  {journal} {American Society
  for Metals, Cleveland}\ }\textbf {\bibinfo {volume} {108}} (\bibinfo {year}
  {1952})}\BibitemShut {NoStop}%
\bibitem [{\citenamefont {Mullins}(1956)}]{mullins1956two}%
  \BibitemOpen
  \bibfield  {author} {\bibinfo {author} {\bibfnamefont {W.~W.}\ \bibnamefont
  {Mullins}},\ }\bibfield  {title} {\bibinfo {title} {Two-dimensional motion of
  idealized grain boundaries},\ }\href {https://doi.org/10.1063/1.1722511}
  {\bibfield  {journal} {\bibinfo  {journal} {J. Appl. Phys.}\ }\textbf
  {\bibinfo {volume} {27}},\ \bibinfo {pages} {900} (\bibinfo {year}
  {1956})}\BibitemShut {NoStop}%
\bibitem [{\citenamefont {MacPherson}\ and\ \citenamefont
  {Srolovitz}(2007)}]{macpherson2007vonneumann}%
  \BibitemOpen
  \bibfield  {author} {\bibinfo {author} {\bibfnamefont {R.~D.}\ \bibnamefont
  {MacPherson}}\ and\ \bibinfo {author} {\bibfnamefont {D.~J.}\ \bibnamefont
  {Srolovitz}},\ }\bibfield  {title} {\bibinfo {title} {The von {Neumann}
  relation generalized to coarsening of three-dimensional microstructures},\
  }\href {https://doi.org/10.1038/nature05745} {\bibfield  {journal} {\bibinfo
  {journal} {Nature}\ }\textbf {\bibinfo {volume} {446}},\ \bibinfo {pages}
  {1053} (\bibinfo {year} {2007})}\BibitemShut {NoStop}%
\bibitem [{\citenamefont {Cahn}\ \emph {et~al.}(2006)\citenamefont {Cahn},
  \citenamefont {Mishin},\ and\ \citenamefont {Suzuki}}]{cahn2006coupling}%
  \BibitemOpen
  \bibfield  {author} {\bibinfo {author} {\bibfnamefont {J.~W.}\ \bibnamefont
  {Cahn}}, \bibinfo {author} {\bibfnamefont {Y.}~\bibnamefont {Mishin}},\ and\
  \bibinfo {author} {\bibfnamefont {A.}~\bibnamefont {Suzuki}},\ }\bibfield
  {title} {\bibinfo {title} {Coupling grain boundary motion to shear
  deformation},\ }\href {https://doi.org/10.1016/j.actamat.2006.08.004}
  {\bibfield  {journal} {\bibinfo  {journal} {Acta Mater.}\ }\textbf {\bibinfo
  {volume} {54}},\ \bibinfo {pages} {4953} (\bibinfo {year}
  {2006})}\BibitemShut {NoStop}%
\bibitem [{\citenamefont {Thomas}\ \emph {et~al.}(2017)\citenamefont {Thomas},
  \citenamefont {Chen}, \citenamefont {Han}, \citenamefont {Purohit},\ and\
  \citenamefont {Srolovitz}}]{thomas2017reconciling}%
  \BibitemOpen
  \bibfield  {author} {\bibinfo {author} {\bibfnamefont {S.~L.}\ \bibnamefont
  {Thomas}}, \bibinfo {author} {\bibfnamefont {K.}~\bibnamefont {Chen}},
  \bibinfo {author} {\bibfnamefont {J.}~\bibnamefont {Han}}, \bibinfo {author}
  {\bibfnamefont {P.~K.}\ \bibnamefont {Purohit}},\ and\ \bibinfo {author}
  {\bibfnamefont {D.~J.}\ \bibnamefont {Srolovitz}},\ }\bibfield  {title}
  {\bibinfo {title} {Reconciling grain growth and shear-coupled grain boundary
  migration},\ }\href {https://doi.org/10.1038/s41467-017-01889-3} {\bibfield
  {journal} {\bibinfo  {journal} {Nat. Commun.}\ }\textbf {\bibinfo {volume}
  {8}},\ \bibinfo {pages} {1} (\bibinfo {year} {2017})}\BibitemShut {NoStop}%
\bibitem [{\citenamefont {Qiu}\ \emph {et~al.}(2025)\citenamefont {Qiu},
  \citenamefont {Srolovitz}, \citenamefont {Rohrer}, \citenamefont {Han},\ and\
  \citenamefont {Salvalaglio}}]{qiu2025why}%
  \BibitemOpen
  \bibfield  {author} {\bibinfo {author} {\bibfnamefont {C.}~\bibnamefont
  {Qiu}}, \bibinfo {author} {\bibfnamefont {D.~J.}\ \bibnamefont {Srolovitz}},
  \bibinfo {author} {\bibfnamefont {G.~S.}\ \bibnamefont {Rohrer}}, \bibinfo
  {author} {\bibfnamefont {J.}~\bibnamefont {Han}},\ and\ \bibinfo {author}
  {\bibfnamefont {M.}~\bibnamefont {Salvalaglio}},\ }\bibfield  {title}
  {\bibinfo {title} {Why grain growth is not curvature flow},\ }\href
  {https://doi.org/10.1073/pnas.2500707122} {\bibfield  {journal} {\bibinfo
  {journal} {Proc. Natl. Acad. Sci. U.S.A}\ }\textbf {\bibinfo {volume}
  {122}},\ \bibinfo {pages} {e2500707122} (\bibinfo {year} {2025})}\BibitemShut
  {NoStop}%
\bibitem [{\citenamefont {Gautier}\ \emph {et~al.}(2025)\citenamefont
  {Gautier}, \citenamefont {Mompiou}, \citenamefont {Renk}, \citenamefont
  {Coupeau}, \citenamefont {Combe}, \citenamefont {Seine},\ and\ \citenamefont
  {Legros}}]{gautier2025quantifying}%
  \BibitemOpen
  \bibfield  {author} {\bibinfo {author} {\bibfnamefont {R.}~\bibnamefont
  {Gautier}}, \bibinfo {author} {\bibfnamefont {F.}~\bibnamefont {Mompiou}},
  \bibinfo {author} {\bibfnamefont {O.}~\bibnamefont {Renk}}, \bibinfo {author}
  {\bibfnamefont {C.}~\bibnamefont {Coupeau}}, \bibinfo {author} {\bibfnamefont
  {N.}~\bibnamefont {Combe}}, \bibinfo {author} {\bibfnamefont
  {G.}~\bibnamefont {Seine}},\ and\ \bibinfo {author} {\bibfnamefont
  {M.}~\bibnamefont {Legros}},\ }\bibfield  {title} {\bibinfo {title}
  {Quantifying grain boundary deformation mechanisms in small-grained metals},\
  }\href {https://doi.org/10.1038/s41586-025-09800-7} {\bibfield  {journal}
  {\bibinfo  {journal} {Nature}\ }\textbf {\bibinfo {volume} {648}},\ \bibinfo
  {pages} {327} (\bibinfo {year} {2025})}\BibitemShut {NoStop}%
\bibitem [{\citenamefont {W{\"o}rner}\ and\ \citenamefont
  {Hazzledine}(1992)}]{worner1992grain}%
  \BibitemOpen
  \bibfield  {author} {\bibinfo {author} {\bibfnamefont {C.}~\bibnamefont
  {W{\"o}rner}}\ and\ \bibinfo {author} {\bibfnamefont {P.}~\bibnamefont
  {Hazzledine}},\ }\bibfield  {title} {\bibinfo {title} {Grain growth
  stagnation by inclusions or pores},\ }\href
  {https://doi.org/10.1007/BF03222320} {\bibfield  {journal} {\bibinfo
  {journal} {Jom}\ }\textbf {\bibinfo {volume} {44}},\ \bibinfo {pages} {16}
  (\bibinfo {year} {1992})}\BibitemShut {NoStop}%
\bibitem [{\citenamefont {Mullins}(1958)}]{mullins1958theeffect}%
  \BibitemOpen
  \bibfield  {author} {\bibinfo {author} {\bibfnamefont {W.}~\bibnamefont
  {Mullins}},\ }\bibfield  {title} {\bibinfo {title} {The effect of thermal
  grooving on grain boundary motion},\ }\href
  {https://doi.org/https://doi.org/10.1016/0001-6160(58)90020-8} {\bibfield
  {journal} {\bibinfo  {journal} {Acta Metall.}\ }\textbf {\bibinfo {volume}
  {6}},\ \bibinfo {pages} {414} (\bibinfo {year} {1958})}\BibitemShut {NoStop}%
\bibitem [{\citenamefont {Bolling}\ and\ \citenamefont
  {Winegard}(1958)}]{bolling1958grain}%
  \BibitemOpen
  \bibfield  {author} {\bibinfo {author} {\bibfnamefont {G.}~\bibnamefont
  {Bolling}}\ and\ \bibinfo {author} {\bibfnamefont {W.}~\bibnamefont
  {Winegard}},\ }\bibfield  {title} {\bibinfo {title} {Grain growth in
  zone-refined lead},\ }\href
  {https://doi.org/https://doi.org/10.1016/0001-6160(58)90148-2} {\bibfield
  {journal} {\bibinfo  {journal} {Acta Metall.}\ }\textbf {\bibinfo {volume}
  {6}},\ \bibinfo {pages} {283} (\bibinfo {year} {1958})}\BibitemShut {NoStop}%
\bibitem [{\citenamefont {Hu}(1974)}]{hu1974grain}%
  \BibitemOpen
  \bibfield  {author} {\bibinfo {author} {\bibfnamefont {H.}~\bibnamefont
  {Hu}},\ }\bibfield  {title} {\bibinfo {title} {Grain growth in zone-refined
  iron},\ }\href {https://doi.org/10.1179/cmq.1974.13.1.275} {\bibfield
  {journal} {\bibinfo  {journal} {Can. Metall. Q.}\ }\textbf {\bibinfo {volume}
  {13}},\ \bibinfo {pages} {275} (\bibinfo {year} {1974})}\BibitemShut
  {NoStop}%
\bibitem [{\citenamefont {Holm}\ and\ \citenamefont
  {Foiles}(2010)}]{holm2010grain}%
  \BibitemOpen
  \bibfield  {author} {\bibinfo {author} {\bibfnamefont {E.~A.}\ \bibnamefont
  {Holm}}\ and\ \bibinfo {author} {\bibfnamefont {S.~M.}\ \bibnamefont
  {Foiles}},\ }\bibfield  {title} {\bibinfo {title} {How grain growth stops: A
  mechanism for grain-growth stagnation in pure materials},\ }\href
  {https://doi.org/10.1126/science.1187833} {\bibfield  {journal} {\bibinfo
  {journal} {Science}\ }\textbf {\bibinfo {volume} {328}},\ \bibinfo {pages}
  {1138} (\bibinfo {year} {2010})}\BibitemShut {NoStop}%
\bibitem [{\citenamefont {Song}\ and\ \citenamefont
  {Deng}(2024)}]{song2024disruptive}%
  \BibitemOpen
  \bibfield  {author} {\bibinfo {author} {\bibfnamefont {X.}~\bibnamefont
  {Song}}\ and\ \bibinfo {author} {\bibfnamefont {C.}~\bibnamefont {Deng}},\
  }\bibfield  {title} {\bibinfo {title} {Disruptive atomic jumps induce grain
  boundary stagnation},\ }\href {https://doi.org/10.1016/j.actamat.2024.120283}
  {\bibfield  {journal} {\bibinfo  {journal} {Acta Mater.}\ }\textbf {\bibinfo
  {volume} {278}},\ \bibinfo {pages} {120283} (\bibinfo {year}
  {2024})}\BibitemShut {NoStop}%
\bibitem [{\citenamefont {Ashby}(1972)}]{ashby1972boundary}%
  \BibitemOpen
  \bibfield  {author} {\bibinfo {author} {\bibfnamefont {M.}~\bibnamefont
  {Ashby}},\ }\bibfield  {title} {\bibinfo {title} {Boundary defects, and
  atomistic aspects of boundary sliding and diffusional creep},\ }\href
  {https://doi.org/10.1016/0039-6028(72)90273-7} {\bibfield  {journal}
  {\bibinfo  {journal} {Surf. Sci.}\ }\textbf {\bibinfo {volume} {31}},\
  \bibinfo {pages} {498} (\bibinfo {year} {1972})}\BibitemShut {NoStop}%
\bibitem [{\citenamefont {Hirth}\ and\ \citenamefont
  {Balluffi}(1973)}]{hirth1973grain}%
  \BibitemOpen
  \bibfield  {author} {\bibinfo {author} {\bibfnamefont {J.~P.}\ \bibnamefont
  {Hirth}}\ and\ \bibinfo {author} {\bibfnamefont {R.~W.}\ \bibnamefont
  {Balluffi}},\ }\bibfield  {title} {\bibinfo {title} {On grain boundary
  dislocations and ledges},\ }\href
  {https://doi.org/10.1016/0001-6160(73)90150-8} {\bibfield  {journal}
  {\bibinfo  {journal} {Acta Metall.}\ }\textbf {\bibinfo {volume} {21}},\
  \bibinfo {pages} {929} (\bibinfo {year} {1973})}\BibitemShut {NoStop}%
\bibitem [{\citenamefont {Qiu}\ \emph {et~al.}(2024)\citenamefont {Qiu},
  \citenamefont {Punke}, \citenamefont {Tian}, \citenamefont {Han},
  \citenamefont {Wang}, \citenamefont {Su}, \citenamefont {Salvalaglio},
  \citenamefont {Pan}, \citenamefont {Srolovitz},\ and\ \citenamefont
  {Han}}]{qiu2024grain}%
  \BibitemOpen
  \bibfield  {author} {\bibinfo {author} {\bibfnamefont {C.}~\bibnamefont
  {Qiu}}, \bibinfo {author} {\bibfnamefont {M.}~\bibnamefont {Punke}}, \bibinfo
  {author} {\bibfnamefont {Y.}~\bibnamefont {Tian}}, \bibinfo {author}
  {\bibfnamefont {Y.}~\bibnamefont {Han}}, \bibinfo {author} {\bibfnamefont
  {S.}~\bibnamefont {Wang}}, \bibinfo {author} {\bibfnamefont {Y.}~\bibnamefont
  {Su}}, \bibinfo {author} {\bibfnamefont {M.}~\bibnamefont {Salvalaglio}},
  \bibinfo {author} {\bibfnamefont {X.}~\bibnamefont {Pan}}, \bibinfo {author}
  {\bibfnamefont {D.~J.}\ \bibnamefont {Srolovitz}},\ and\ \bibinfo {author}
  {\bibfnamefont {J.}~\bibnamefont {Han}},\ }\bibfield  {title} {\bibinfo
  {title} {Grain boundaries are brownian ratchets},\ }\href
  {https://doi.org/10.1126/science.adp1516} {\bibfield  {journal} {\bibinfo
  {journal} {Science}\ }\textbf {\bibinfo {volume} {385}},\ \bibinfo {pages}
  {980} (\bibinfo {year} {2024})}\BibitemShut {NoStop}%
\bibitem [{\citenamefont {Han}\ \emph {et~al.}(2018)\citenamefont {Han},
  \citenamefont {Thomas},\ and\ \citenamefont {Srolovitz}}]{han2018grain}%
  \BibitemOpen
  \bibfield  {author} {\bibinfo {author} {\bibfnamefont {J.}~\bibnamefont
  {Han}}, \bibinfo {author} {\bibfnamefont {S.~L.}\ \bibnamefont {Thomas}},\
  and\ \bibinfo {author} {\bibfnamefont {D.~J.}\ \bibnamefont {Srolovitz}},\
  }\bibfield  {title} {\bibinfo {title} {{Grain-boundary kinetics: A unified
  approach}},\ }\href {https://doi.org/10.1016/j.pmatsci.2018.05.004}
  {\bibfield  {journal} {\bibinfo  {journal} {Prog. Mater. Sci.}\ }\textbf
  {\bibinfo {volume} {98}},\ \bibinfo {pages} {386} (\bibinfo {year}
  {2018})}\BibitemShut {NoStop}%
\bibitem [{\citenamefont {Wei}\ \emph {et~al.}(2021)\citenamefont {Wei},
  \citenamefont {Feng}, \citenamefont {Ishikawa}, \citenamefont {Yokoi},
  \citenamefont {Matsunaga}, \citenamefont {Shibata},\ and\ \citenamefont
  {Ikuhara}}]{wei2021direct}%
  \BibitemOpen
  \bibfield  {author} {\bibinfo {author} {\bibfnamefont {J.}~\bibnamefont
  {Wei}}, \bibinfo {author} {\bibfnamefont {B.}~\bibnamefont {Feng}}, \bibinfo
  {author} {\bibfnamefont {R.}~\bibnamefont {Ishikawa}}, \bibinfo {author}
  {\bibfnamefont {T.}~\bibnamefont {Yokoi}}, \bibinfo {author} {\bibfnamefont
  {K.}~\bibnamefont {Matsunaga}}, \bibinfo {author} {\bibfnamefont
  {N.}~\bibnamefont {Shibata}},\ and\ \bibinfo {author} {\bibfnamefont
  {Y.}~\bibnamefont {Ikuhara}},\ }\bibfield  {title} {\bibinfo {title} {Direct
  imaging of atomistic grain boundary migration},\ }\href
  {https://doi.org/https://doi.org/10.1038/s41563-020-00879-z} {\bibfield
  {journal} {\bibinfo  {journal} {Nat. Mater.}\ }\textbf {\bibinfo {volume}
  {20}},\ \bibinfo {pages} {951} (\bibinfo {year} {2021})}\BibitemShut
  {NoStop}%
\bibitem [{\citenamefont {Wei}\ \emph {et~al.}(2022)\citenamefont {Wei},
  \citenamefont {Feng}, \citenamefont {Tochigi}, \citenamefont {Shibata},\ and\
  \citenamefont {Ikuhara}}]{wei2022direct}%
  \BibitemOpen
  \bibfield  {author} {\bibinfo {author} {\bibfnamefont {J.}~\bibnamefont
  {Wei}}, \bibinfo {author} {\bibfnamefont {B.}~\bibnamefont {Feng}}, \bibinfo
  {author} {\bibfnamefont {E.}~\bibnamefont {Tochigi}}, \bibinfo {author}
  {\bibfnamefont {N.}~\bibnamefont {Shibata}},\ and\ \bibinfo {author}
  {\bibfnamefont {Y.}~\bibnamefont {Ikuhara}},\ }\bibfield  {title} {\bibinfo
  {title} {Direct imaging of the disconnection climb mediated point defects
  absorption by a grain boundary},\ }\href
  {https://doi.org/https://doi.org/10.1038/s41467-022-29162-2} {\bibfield
  {journal} {\bibinfo  {journal} {Nat. Commun.}\ }\textbf {\bibinfo {volume}
  {13}},\ \bibinfo {pages} {1455} (\bibinfo {year} {2022})}\BibitemShut
  {NoStop}%
\bibitem [{\citenamefont {Feng}\ \emph {et~al.}(2023)\citenamefont {Feng},
  \citenamefont {Wei}, \citenamefont {Shibata},\ and\ \citenamefont
  {Ikuhara}}]{feng2023atomistic}%
  \BibitemOpen
  \bibfield  {author} {\bibinfo {author} {\bibfnamefont {B.}~\bibnamefont
  {Feng}}, \bibinfo {author} {\bibfnamefont {J.}~\bibnamefont {Wei}}, \bibinfo
  {author} {\bibfnamefont {N.}~\bibnamefont {Shibata}},\ and\ \bibinfo {author}
  {\bibfnamefont {Y.}~\bibnamefont {Ikuhara}},\ }\bibfield  {title} {\bibinfo
  {title} {Atomistic grain boundary migration in al2o3},\ }\href
  {https://doi.org/https://doi.org/10.1002/ces2.10169} {\bibfield  {journal}
  {\bibinfo  {journal} {Int. J. Ceram. Eng. Sci.}\ }\textbf {\bibinfo {volume}
  {5}},\ \bibinfo {pages} {e10169} (\bibinfo {year} {2023})}\BibitemShut
  {NoStop}%
\bibitem [{\citenamefont {Han}\ \emph {et~al.}(2016)\citenamefont {Han},
  \citenamefont {Vitek},\ and\ \citenamefont {Srolovitz}}]{han2016grain}%
  \BibitemOpen
  \bibfield  {author} {\bibinfo {author} {\bibfnamefont {J.}~\bibnamefont
  {Han}}, \bibinfo {author} {\bibfnamefont {V.}~\bibnamefont {Vitek}},\ and\
  \bibinfo {author} {\bibfnamefont {D.~J.}\ \bibnamefont {Srolovitz}},\
  }\bibfield  {title} {\bibinfo {title} {Grain-boundary metastability and its
  statistical properties},\ }\href
  {https://doi.org/https://doi.org/10.1016/j.actamat.2015.11.035} {\bibfield
  {journal} {\bibinfo  {journal} {Acta Mater.}\ }\textbf {\bibinfo {volume}
  {104}},\ \bibinfo {pages} {259} (\bibinfo {year} {2016})}\BibitemShut
  {NoStop}%
\bibitem [{\citenamefont {Cantwell}\ \emph {et~al.}(2020)\citenamefont
  {Cantwell}, \citenamefont {Frolov}, \citenamefont {Rupert}, \citenamefont
  {Krause}, \citenamefont {Marvel}, \citenamefont {Rohrer}, \citenamefont
  {Rickman},\ and\ \citenamefont {Harmer}}]{Cantwell2020}%
  \BibitemOpen
  \bibfield  {author} {\bibinfo {author} {\bibfnamefont {P.~R.}\ \bibnamefont
  {Cantwell}}, \bibinfo {author} {\bibfnamefont {T.}~\bibnamefont {Frolov}},
  \bibinfo {author} {\bibfnamefont {T.~J.}\ \bibnamefont {Rupert}}, \bibinfo
  {author} {\bibfnamefont {A.~R.}\ \bibnamefont {Krause}}, \bibinfo {author}
  {\bibfnamefont {C.~J.}\ \bibnamefont {Marvel}}, \bibinfo {author}
  {\bibfnamefont {G.~S.}\ \bibnamefont {Rohrer}}, \bibinfo {author}
  {\bibfnamefont {J.~M.}\ \bibnamefont {Rickman}},\ and\ \bibinfo {author}
  {\bibfnamefont {M.~P.}\ \bibnamefont {Harmer}},\ }\bibfield  {title}
  {\bibinfo {title} {Grain boundary complexion transitions},\ }\href
  {https://doi.org/https://doi.org/10.1146/annurev-matsci-081619-114055}
  {\bibfield  {journal} {\bibinfo  {journal} {Annu. Rev. Mater. Res.}\ }\textbf
  {\bibinfo {volume} {50}},\ \bibinfo {pages} {465} (\bibinfo {year}
  {2020})}\BibitemShut {NoStop}%
\bibitem [{\citenamefont {Skogvoll}\ \emph {et~al.}(2021)\citenamefont
  {Skogvoll}, \citenamefont {Skaugen}, \citenamefont {Angheluta},\ and\
  \citenamefont {Vi{\~n}als}}]{skogvoll2021dislocation}%
  \BibitemOpen
  \bibfield  {author} {\bibinfo {author} {\bibfnamefont {V.}~\bibnamefont
  {Skogvoll}}, \bibinfo {author} {\bibfnamefont {A.}~\bibnamefont {Skaugen}},
  \bibinfo {author} {\bibfnamefont {L.}~\bibnamefont {Angheluta}},\ and\
  \bibinfo {author} {\bibfnamefont {J.}~\bibnamefont {Vi{\~n}als}},\ }\bibfield
   {title} {\bibinfo {title} {Dislocation nucleation in the phase-field crystal
  model},\ }\href {https://doi.org/10.1103/PhysRevB.103.014107} {\bibfield
  {journal} {\bibinfo  {journal} {Phys. Rev. B}\ }\textbf {\bibinfo {volume}
  {103}},\ \bibinfo {pages} {014107} (\bibinfo {year} {2021})}\BibitemShut
  {NoStop}%
\bibitem [{\citenamefont {Stukowski}(2009)}]{stukowski2009visualization}%
  \BibitemOpen
  \bibfield  {author} {\bibinfo {author} {\bibfnamefont {A.}~\bibnamefont
  {Stukowski}},\ }\bibfield  {title} {\bibinfo {title} {Visualization and
  analysis of atomistic simulation data with ovito--the open visualization
  tool},\ }\href {https://doi.org/10.1088/0965-0393/18/1/015012} {\bibfield
  {journal} {\bibinfo  {journal} {Model. Simul. Mater. Sci. Eng.}\ }\textbf
  {\bibinfo {volume} {18}},\ \bibinfo {pages} {015012} (\bibinfo {year}
  {2009})}\BibitemShut {NoStop}%
\bibitem [{\citenamefont {Elder}\ \emph {et~al.}(2002)\citenamefont {Elder},
  \citenamefont {Katakowski}, \citenamefont {Haataja},\ and\ \citenamefont
  {Grant}}]{Elder2002}%
  \BibitemOpen
  \bibfield  {author} {\bibinfo {author} {\bibfnamefont {K.~R.}\ \bibnamefont
  {Elder}}, \bibinfo {author} {\bibfnamefont {M.}~\bibnamefont {Katakowski}},
  \bibinfo {author} {\bibfnamefont {M.}~\bibnamefont {Haataja}},\ and\ \bibinfo
  {author} {\bibfnamefont {M.}~\bibnamefont {Grant}},\ }\bibfield  {title}
  {\bibinfo {title} {{Modeling Elasticity in Crystal Growth}},\ }\href
  {https://doi.org/https://doi.org/10.1103/PhysRevLett.88.245701} {\bibfield
  {journal} {\bibinfo  {journal} {Phys. Rev. Lett.}\ }\textbf {\bibinfo
  {volume} {88}},\ \bibinfo {pages} {245701} (\bibinfo {year}
  {2002})}\BibitemShut {NoStop}%
\bibitem [{\citenamefont {Trautt}\ \emph {et~al.}(2012)\citenamefont {Trautt},
  \citenamefont {Adland}, \citenamefont {Karma},\ and\ \citenamefont
  {Mishin}}]{trautt2012coupled}%
  \BibitemOpen
  \bibfield  {author} {\bibinfo {author} {\bibfnamefont {Z.}~\bibnamefont
  {Trautt}}, \bibinfo {author} {\bibfnamefont {A.}~\bibnamefont {Adland}},
  \bibinfo {author} {\bibfnamefont {A.}~\bibnamefont {Karma}},\ and\ \bibinfo
  {author} {\bibfnamefont {Y.}~\bibnamefont {Mishin}},\ }\bibfield  {title}
  {\bibinfo {title} {{Coupled motion of asymmetrical tilt grain boundaries:
  Molecular dynamics and phase field crystal simulations}},\ }\href
  {https://doi.org/10.1016/j.actamat.2012.08.018} {\bibfield  {journal}
  {\bibinfo  {journal} {Acta Mater.}\ }\textbf {\bibinfo {volume} {60}},\
  \bibinfo {pages} {6528} (\bibinfo {year} {2012})}\BibitemShut {NoStop}%
\bibitem [{\citenamefont {Berry}\ \emph {et~al.}(2014)\citenamefont {Berry},
  \citenamefont {Provatas}, \citenamefont {Rottler},\ and\ \citenamefont
  {Sinclair}}]{Berry2014}%
  \BibitemOpen
  \bibfield  {author} {\bibinfo {author} {\bibfnamefont {J.}~\bibnamefont
  {Berry}}, \bibinfo {author} {\bibfnamefont {N.}~\bibnamefont {Provatas}},
  \bibinfo {author} {\bibfnamefont {J.}~\bibnamefont {Rottler}},\ and\ \bibinfo
  {author} {\bibfnamefont {C.~W.}\ \bibnamefont {Sinclair}},\ }\bibfield
  {title} {\bibinfo {title} {Phase field crystal modeling as a unified
  atomistic approach to defect dynamics},\ }\href
  {https://doi.org/https://doi.org/10.1103/PhysRevB.89.214117} {\bibfield
  {journal} {\bibinfo  {journal} {Phys. Rev. B}\ }\textbf {\bibinfo {volume}
  {89}},\ \bibinfo {pages} {214117} (\bibinfo {year} {2014})}\BibitemShut
  {NoStop}%
\bibitem [{\citenamefont {Skogvoll}\ \emph
  {et~al.}(2022{\natexlab{a}})\citenamefont {Skogvoll}, \citenamefont
  {Angheluta}, \citenamefont {Skaugen}, \citenamefont {Salvalaglio},\ and\
  \citenamefont {Vi{\~n}als}}]{skogvoll2022phase}%
  \BibitemOpen
  \bibfield  {author} {\bibinfo {author} {\bibfnamefont {V.}~\bibnamefont
  {Skogvoll}}, \bibinfo {author} {\bibfnamefont {L.}~\bibnamefont {Angheluta}},
  \bibinfo {author} {\bibfnamefont {A.}~\bibnamefont {Skaugen}}, \bibinfo
  {author} {\bibfnamefont {M.}~\bibnamefont {Salvalaglio}},\ and\ \bibinfo
  {author} {\bibfnamefont {J.}~\bibnamefont {Vi{\~n}als}},\ }\bibfield  {title}
  {\bibinfo {title} {A phase field crystal theory of the kinematics of
  dislocation lines},\ }\href
  {https://doi.org/https://doi.org/10.1016/j.jmps.2022.104932} {\bibfield
  {journal} {\bibinfo  {journal} {J. Mech. Phys. Solids}\ }\textbf {\bibinfo
  {volume} {166}},\ \bibinfo {pages} {104932} (\bibinfo {year}
  {2022}{\natexlab{a}})}\BibitemShut {NoStop}%
\bibitem [{\citenamefont {Bjerre}\ \emph {et~al.}(2013)\citenamefont {Bjerre},
  \citenamefont {Tarp}, \citenamefont {Angheluta},\ and\ \citenamefont
  {Mathiesen}}]{Bjerre2013}%
  \BibitemOpen
  \bibfield  {author} {\bibinfo {author} {\bibfnamefont {M.}~\bibnamefont
  {Bjerre}}, \bibinfo {author} {\bibfnamefont {J.~M.}\ \bibnamefont {Tarp}},
  \bibinfo {author} {\bibfnamefont {L.}~\bibnamefont {Angheluta}},\ and\
  \bibinfo {author} {\bibfnamefont {J.}~\bibnamefont {Mathiesen}},\ }\bibfield
  {title} {\bibinfo {title} {Rotation-induced grain growth and stagnation in
  phase-field crystal models},\ }\href
  {https://doi.org/10.1103/PhysRevE.88.020401} {\bibfield  {journal} {\bibinfo
  {journal} {Phys. Rev. E}\ }\textbf {\bibinfo {volume} {88}},\ \bibinfo
  {pages} {020401} (\bibinfo {year} {2013})}\BibitemShut {NoStop}%
\bibitem [{\citenamefont {Backofen}\ \emph {et~al.}(2014)\citenamefont
  {Backofen}, \citenamefont {Barmak}, \citenamefont {Elder},\ and\
  \citenamefont {Voigt}}]{backofen2014capturing}%
  \BibitemOpen
  \bibfield  {author} {\bibinfo {author} {\bibfnamefont {R.}~\bibnamefont
  {Backofen}}, \bibinfo {author} {\bibfnamefont {K.}~\bibnamefont {Barmak}},
  \bibinfo {author} {\bibfnamefont {K.}~\bibnamefont {Elder}},\ and\ \bibinfo
  {author} {\bibfnamefont {A.}~\bibnamefont {Voigt}},\ }\bibfield  {title}
  {\bibinfo {title} {Capturing the complex physics behind universal grain size
  distributions in thin metallic films},\ }\href
  {https://doi.org/https://doi.org/10.1016/j.actamat.2013.11.034} {\bibfield
  {journal} {\bibinfo  {journal} {Acta Mater.}\ }\textbf {\bibinfo {volume}
  {64}},\ \bibinfo {pages} {72} (\bibinfo {year} {2014})}\BibitemShut {NoStop}%
\bibitem [{\citenamefont {Heinonen}\ \emph {et~al.}(2016)\citenamefont
  {Heinonen}, \citenamefont {Achim}, \citenamefont {Kosterlitz}, \citenamefont
  {Ying}, \citenamefont {Lowengrub},\ and\ \citenamefont
  {Ala-Nissila}}]{HeinonenPRL2016}%
  \BibitemOpen
  \bibfield  {author} {\bibinfo {author} {\bibfnamefont {V.}~\bibnamefont
  {Heinonen}}, \bibinfo {author} {\bibfnamefont {C.~V.}\ \bibnamefont {Achim}},
  \bibinfo {author} {\bibfnamefont {J.~M.}\ \bibnamefont {Kosterlitz}},
  \bibinfo {author} {\bibfnamefont {S.-C.}\ \bibnamefont {Ying}}, \bibinfo
  {author} {\bibfnamefont {J.}~\bibnamefont {Lowengrub}},\ and\ \bibinfo
  {author} {\bibfnamefont {T.}~\bibnamefont {Ala-Nissila}},\ }\bibfield
  {title} {\bibinfo {title} {{Consistent Hydrodynamics for Phase Field
  Crystals}},\ }\href
  {https://doi.org/https://doi.org/10.1103/PhysRevLett.116.024303} {\bibfield
  {journal} {\bibinfo  {journal} {Phys. Rev. Lett.}\ }\textbf {\bibinfo
  {volume} {116}},\ \bibinfo {pages} {024303} (\bibinfo {year}
  {2016})}\BibitemShut {NoStop}%
\bibitem [{\citenamefont {Skogvoll}\ \emph
  {et~al.}(2022{\natexlab{b}})\citenamefont {Skogvoll}, \citenamefont
  {Salvalaglio},\ and\ \citenamefont {Angheluta}}]{skogvoll2022hydrodynamic}%
  \BibitemOpen
  \bibfield  {author} {\bibinfo {author} {\bibfnamefont {V.}~\bibnamefont
  {Skogvoll}}, \bibinfo {author} {\bibfnamefont {M.}~\bibnamefont
  {Salvalaglio}},\ and\ \bibinfo {author} {\bibfnamefont {L.}~\bibnamefont
  {Angheluta}},\ }\bibfield  {title} {\bibinfo {title} {Hydrodynamic phase
  field crystal approach to interfaces, dislocations, and multi-grain
  networks},\ }\href {https://doi.org/10.1088/1361-651X/ac9493} {\bibfield
  {journal} {\bibinfo  {journal} {Model. Simul. Mater. Sci. Eng.}\ }\textbf
  {\bibinfo {volume} {30}},\ \bibinfo {pages} {084002} (\bibinfo {year}
  {2022}{\natexlab{b}})}\BibitemShut {NoStop}%
\bibitem [{Sup()}]{SupplementalMaterial}%
  \BibitemOpen
  \href@noop {} {}\bibinfo {note} {Supplemental material includes technical
  details concerning: (i) Local averaging; (ii) Numerical methods for PFC
  simulations; (iii) Grain counting method.}\BibitemShut {Stop}%
\bibitem [{\citenamefont {Emmerich}\ \emph {et~al.}(2012)\citenamefont
  {Emmerich}, \citenamefont {L{\"{o}}wen}, \citenamefont {Wittkowski},
  \citenamefont {Gruhn}, \citenamefont {T{\'{o}}th}, \citenamefont {Tegze},\
  and\ \citenamefont {Gr{\'{a}}n{\'{a}}sy}}]{Emmerich2012}%
  \BibitemOpen
  \bibfield  {author} {\bibinfo {author} {\bibfnamefont {H.}~\bibnamefont
  {Emmerich}}, \bibinfo {author} {\bibfnamefont {H.}~\bibnamefont
  {L{\"{o}}wen}}, \bibinfo {author} {\bibfnamefont {R.}~\bibnamefont
  {Wittkowski}}, \bibinfo {author} {\bibfnamefont {T.}~\bibnamefont {Gruhn}},
  \bibinfo {author} {\bibfnamefont {G.~I.}\ \bibnamefont {T{\'{o}}th}},
  \bibinfo {author} {\bibfnamefont {G.}~\bibnamefont {Tegze}},\ and\ \bibinfo
  {author} {\bibfnamefont {L.}~\bibnamefont {Gr{\'{a}}n{\'{a}}sy}},\ }\bibfield
   {title} {\bibinfo {title} {{Phase-field-crystal models for condensed matter
  dynamics on atomic length and diffusive time scales: an overview}},\ }\href
  {https://doi.org/https://doi.org/10.1080/00018732.2012.737555} {\bibfield
  {journal} {\bibinfo  {journal} {Adv. Phys.}\ }\textbf {\bibinfo {volume}
  {61}},\ \bibinfo {pages} {665} (\bibinfo {year} {2012})}\BibitemShut
  {NoStop}%
\bibitem [{Note1()}]{Note1}%
  \BibitemOpen
  \bibinfo {note} {The specific functional form follows as the simplest even
  term penalizing gradient, in analogy with classical phase field
  models}\BibitemShut {NoStop}%
\bibitem [{\citenamefont {Estrin}\ and\ \citenamefont
  {Lücke}(1982)}]{estrin1982theory}%
  \BibitemOpen
  \bibfield  {author} {\bibinfo {author} {\bibfnamefont {Y.}~\bibnamefont
  {Estrin}}\ and\ \bibinfo {author} {\bibfnamefont {K.}~\bibnamefont
  {Lücke}},\ }\bibfield  {title} {\bibinfo {title} {Theory of
  vacancy-controlled grain boundary motion},\ }\href
  {https://doi.org/https://doi.org/10.1016/0001-6160(82)90206-1} {\bibfield
  {journal} {\bibinfo  {journal} {Acta Metall.}\ }\textbf {\bibinfo {volume}
  {30}},\ \bibinfo {pages} {983} (\bibinfo {year} {1982})}\BibitemShut
  {NoStop}%
\bibitem [{\citenamefont {McFadden}\ \emph {et~al.}(2020)\citenamefont
  {McFadden}, \citenamefont {Boettinger},\ and\ \citenamefont
  {Mishin}}]{McFadden2020}%
  \BibitemOpen
  \bibfield  {author} {\bibinfo {author} {\bibfnamefont {G.}~\bibnamefont
  {McFadden}}, \bibinfo {author} {\bibfnamefont {W.}~\bibnamefont
  {Boettinger}},\ and\ \bibinfo {author} {\bibfnamefont {Y.}~\bibnamefont
  {Mishin}},\ }\bibfield  {title} {\bibinfo {title} {Effect of vacancy creation
  and annihilation on grain boundary motion},\ }\href
  {https://doi.org/https://doi.org/10.1016/j.actamat.2019.11.044} {\bibfield
  {journal} {\bibinfo  {journal} {Acta Mater.}\ }\textbf {\bibinfo {volume}
  {185}},\ \bibinfo {pages} {66} (\bibinfo {year} {2020})}\BibitemShut
  {NoStop}%
\bibitem [{\citenamefont {Guo}\ \emph {et~al.}(2016)\citenamefont {Guo},
  \citenamefont {Wang}, \citenamefont {Wang}, \citenamefont {Li}, \citenamefont
  {Guo},\ and\ \citenamefont {Huang}}]{guo2016interfacial}%
  \BibitemOpen
  \bibfield  {author} {\bibinfo {author} {\bibfnamefont {C.}~\bibnamefont
  {Guo}}, \bibinfo {author} {\bibfnamefont {J.}~\bibnamefont {Wang}}, \bibinfo
  {author} {\bibfnamefont {Z.}~\bibnamefont {Wang}}, \bibinfo {author}
  {\bibfnamefont {J.}~\bibnamefont {Li}}, \bibinfo {author} {\bibfnamefont
  {Y.}~\bibnamefont {Guo}},\ and\ \bibinfo {author} {\bibfnamefont
  {Y.}~\bibnamefont {Huang}},\ }\bibfield  {title} {\bibinfo {title}
  {Interfacial free energy adjustable phase field crystal model for homogeneous
  nucleation},\ }\href {https://doi.org/10.1039/C6SM00774K} {\bibfield
  {journal} {\bibinfo  {journal} {Soft Matter}\ }\textbf {\bibinfo {volume}
  {12}},\ \bibinfo {pages} {4666} (\bibinfo {year} {2016})}\BibitemShut
  {NoStop}%
\bibitem [{\citenamefont {Salvalaglio}\ \emph {et~al.}(2017)\citenamefont
  {Salvalaglio}, \citenamefont {Backofen}, \citenamefont {Voigt},\ and\
  \citenamefont {Elder}}]{salvalaglio2017controlling}%
  \BibitemOpen
  \bibfield  {author} {\bibinfo {author} {\bibfnamefont {M.}~\bibnamefont
  {Salvalaglio}}, \bibinfo {author} {\bibfnamefont {R.}~\bibnamefont
  {Backofen}}, \bibinfo {author} {\bibfnamefont {A.}~\bibnamefont {Voigt}},\
  and\ \bibinfo {author} {\bibfnamefont {K.~R.}\ \bibnamefont {Elder}},\
  }\bibfield  {title} {\bibinfo {title} {Controlling the energy of defects and
  interfaces in the amplitude expansion of the phase-field crystal model},\
  }\href {https://doi.org/https://doi.org/10.1103/PhysRevE.96.023301}
  {\bibfield  {journal} {\bibinfo  {journal} {Phys. Rev. E}\ }\textbf {\bibinfo
  {volume} {96}},\ \bibinfo {pages} {023301} (\bibinfo {year}
  {2017})}\BibitemShut {NoStop}%
\bibitem [{\citenamefont {Coelho}\ \emph {et~al.}(2024)\citenamefont {Coelho},
  \citenamefont {Burns}, \citenamefont {Wilson},\ and\ \citenamefont
  {Provatas}}]{coelho2024generalizing}%
  \BibitemOpen
  \bibfield  {author} {\bibinfo {author} {\bibfnamefont {D.~L.}\ \bibnamefont
  {Coelho}}, \bibinfo {author} {\bibfnamefont {D.}~\bibnamefont {Burns}},
  \bibinfo {author} {\bibfnamefont {E.}~\bibnamefont {Wilson}},\ and\ \bibinfo
  {author} {\bibfnamefont {N.}~\bibnamefont {Provatas}},\ }\bibfield  {title}
  {\bibinfo {title} {Generalizing the structural phase field crystal approach
  for modeling solid-liquid-vapor phase transformations in pure materials},\
  }\href {https://doi.org/10.1103/PhysRevMaterials.8.093402} {\bibfield
  {journal} {\bibinfo  {journal} {Phys. Rev. Mater.}\ }\textbf {\bibinfo
  {volume} {8}},\ \bibinfo {pages} {093402} (\bibinfo {year}
  {2024})}\BibitemShut {NoStop}%
\bibitem [{\citenamefont {Elder}\ and\ \citenamefont
  {Grant}(2004)}]{Elder2004}%
  \BibitemOpen
  \bibfield  {author} {\bibinfo {author} {\bibfnamefont {K.~R.}\ \bibnamefont
  {Elder}}\ and\ \bibinfo {author} {\bibfnamefont {M.}~\bibnamefont {Grant}},\
  }\bibfield  {title} {\bibinfo {title} {{Modeling elastic and plastic
  deformations in nonequilibrium processing using phase field crystals}},\
  }\href {https://doi.org/https://doi.org/10.1103/PhysRevE.70.051605}
  {\bibfield  {journal} {\bibinfo  {journal} {Phys. Rev. E}\ }\textbf {\bibinfo
  {volume} {70}},\ \bibinfo {pages} {051605} (\bibinfo {year}
  {2004})}\BibitemShut {NoStop}%
\bibitem [{\citenamefont {van Teeffelen}\ \emph {et~al.}(2013)\citenamefont
  {van Teeffelen}, \citenamefont {Achim},\ and\ \citenamefont
  {L{\"o}wen}}]{van2013vacancy}%
  \BibitemOpen
  \bibfield  {author} {\bibinfo {author} {\bibfnamefont {S.}~\bibnamefont {van
  Teeffelen}}, \bibinfo {author} {\bibfnamefont {C.~V.}\ \bibnamefont
  {Achim}},\ and\ \bibinfo {author} {\bibfnamefont {H.}~\bibnamefont
  {L{\"o}wen}},\ }\bibfield  {title} {\bibinfo {title} {Vacancy diffusion in
  colloidal crystals as determined by dynamical density-functional theory and
  the phase-field-crystal model},\ }\href
  {https://doi.org/https://doi.org/10.1103/PhysRevE.87.022306} {\bibfield
  {journal} {\bibinfo  {journal} {Phys. Rev. E}\ }\textbf {\bibinfo {volume}
  {87}},\ \bibinfo {pages} {022306} (\bibinfo {year} {2013})}\BibitemShut
  {NoStop}%
\bibitem [{\citenamefont {Edelsbrunner}\ and\ \citenamefont
  {Harer}(2010)}]{edelsbrunner2022computational}%
  \BibitemOpen
  \bibfield  {author} {\bibinfo {author} {\bibfnamefont {H.}~\bibnamefont
  {Edelsbrunner}}\ and\ \bibinfo {author} {\bibfnamefont {J.~L.}\ \bibnamefont
  {Harer}},\ }\href@noop {} {\emph {\bibinfo {title}
  {\href{https://bookstore.ams.org/mbk-69}{Computational topology: an
  introduction}}}}\ (\bibinfo  {publisher} {American Mathematical Society},\
  \bibinfo {year} {2010})\BibitemShut {NoStop}%
\bibitem [{\citenamefont {Anderson}\ \emph {et~al.}(2017)\citenamefont
  {Anderson}, \citenamefont {Hirth},\ and\ \citenamefont
  {Lothe}}]{anderson2017}%
  \BibitemOpen
  \bibfield  {author} {\bibinfo {author} {\bibfnamefont {P.}~\bibnamefont
  {Anderson}}, \bibinfo {author} {\bibfnamefont {J.}~\bibnamefont {Hirth}},\
  and\ \bibinfo {author} {\bibfnamefont {J.}~\bibnamefont {Lothe}},\
  }\href@noop {} {\emph {\bibinfo {title}
  {\href{https://www.cambridge.org/core/books/abs/theory-of-dislocations/contents/C0C73A76837EE45A292254BCA4F5732C}{Theory
  of Dislocations}}}}\ (\bibinfo  {publisher} {Cambridge University Press},\
  \bibinfo {year} {2017})\BibitemShut {NoStop}%
\bibitem [{\citenamefont {Sutton}\ and\ \citenamefont
  {Balluffi}(1995)}]{SuttonBalluffi}%
  \BibitemOpen
  \bibfield  {author} {\bibinfo {author} {\bibfnamefont {A.~P.}\ \bibnamefont
  {Sutton}}\ and\ \bibinfo {author} {\bibfnamefont {R.~W.}\ \bibnamefont
  {Balluffi}},\ }\href@noop {} {\emph {\bibinfo {title}
  {\href{https://global.oup.com/academic/product/interfaces-in-crystalline-materials-9780198500612}{Interfaces
  in Crystalline Materials}}}}\ (\bibinfo  {publisher} {Oxford University
  Press},\ \bibinfo {year} {1995})\BibitemShut {NoStop}%
\bibitem [{\citenamefont {Zhang}\ \emph {et~al.}(2020)\citenamefont {Zhang},
  \citenamefont {Ludwig}, \citenamefont {Zhang}, \citenamefont {Sørensen},
  \citenamefont {Rowenhorst}, \citenamefont {Yamanaka}, \citenamefont
  {Voorhees},\ and\ \citenamefont {Poulsen}}]{zhang2020grain}%
  \BibitemOpen
  \bibfield  {author} {\bibinfo {author} {\bibfnamefont {J.}~\bibnamefont
  {Zhang}}, \bibinfo {author} {\bibfnamefont {W.}~\bibnamefont {Ludwig}},
  \bibinfo {author} {\bibfnamefont {Y.}~\bibnamefont {Zhang}}, \bibinfo
  {author} {\bibfnamefont {H.~H.~B.}\ \bibnamefont {Sørensen}}, \bibinfo
  {author} {\bibfnamefont {D.~J.}\ \bibnamefont {Rowenhorst}}, \bibinfo
  {author} {\bibfnamefont {A.}~\bibnamefont {Yamanaka}}, \bibinfo {author}
  {\bibfnamefont {P.~W.}\ \bibnamefont {Voorhees}},\ and\ \bibinfo {author}
  {\bibfnamefont {H.~F.}\ \bibnamefont {Poulsen}},\ }\bibfield  {title}
  {\bibinfo {title} {Grain boundary mobilities in polycrystals},\ }\href
  {https://doi.org/https://doi.org/10.1016/j.actamat.2020.03.044} {\bibfield
  {journal} {\bibinfo  {journal} {Acta Mater.}\ }\textbf {\bibinfo {volume}
  {191}},\ \bibinfo {pages} {211} (\bibinfo {year} {2020})}\BibitemShut
  {NoStop}%
\bibitem [{\citenamefont {Xu}\ \emph {et~al.}(2026)\citenamefont {Xu},
  \citenamefont {Abdeljawad},\ and\ \citenamefont {Rohrer}}]{xu2026scripta}%
  \BibitemOpen
  \bibfield  {author} {\bibinfo {author} {\bibfnamefont {Z.}~\bibnamefont
  {Xu}}, \bibinfo {author} {\bibfnamefont {F.}~\bibnamefont {Abdeljawad}},\
  and\ \bibinfo {author} {\bibfnamefont {G.~S.}\ \bibnamefont {Rohrer}},\
  }\bibfield  {title} {\bibinfo {title} {Can the grain boundary stiffness
  driving force explain observed grain boundary migration rates in
  polycrystals?},\ }\href
  {https://doi.org/https://doi.org/10.1016/j.scriptamat.2025.117135} {\bibfield
   {journal} {\bibinfo  {journal} {Scripta Mater.}\ }\textbf {\bibinfo {volume}
  {274}},\ \bibinfo {pages} {117135} (\bibinfo {year} {2026})}\BibitemShut
  {NoStop}%
\end{thebibliography}%


\begin{thebibliography}{19}%
\makeatletter
\providecommand \@ifxundefined [1]{%
 \@ifx{#1\undefined}
}%
\providecommand \@ifnum [1]{%
 \ifnum #1\expandafter \@firstoftwo
 \else \expandafter \@secondoftwo
 \fi
}%
\providecommand \@ifx [1]{%
 \ifx #1\expandafter \@firstoftwo
 \else \expandafter \@secondoftwo
 \fi
}%
\providecommand \natexlab [1]{#1}%
\providecommand \enquote  [1]{``#1''}%
\providecommand \bibnamefont  [1]{#1}%
\providecommand \bibfnamefont [1]{#1}%
\providecommand \citenamefont [1]{#1}%
\providecommand \href@noop [0]{\@secondoftwo}%
\providecommand \href [0]{\begingroup \@sanitize@url \@href}%
\providecommand \@href[1]{\@@startlink{#1}\@@href}%
\providecommand \@@href[1]{\endgroup#1\@@endlink}%
\providecommand \@sanitize@url [0]{\catcode `\\12\catcode `\$12\catcode
  `\&12\catcode `\#12\catcode `\^12\catcode `\_12\catcode `\%12\relax}%
\providecommand \@@startlink[1]{}%
\providecommand \@@endlink[0]{}%
\providecommand \url  [0]{\begingroup\@sanitize@url \@url }%
\providecommand \@url [1]{\endgroup\@href {#1}{\urlprefix }}%
\providecommand \urlprefix  [0]{URL }%
\providecommand \Eprint [0]{\href }%
\providecommand \doibase [0]{http://dx.doi.org/}%
\providecommand \selectlanguage [0]{\@gobble}%
\providecommand \bibinfo  [0]{\@secondoftwo}%
\providecommand \bibfield  [0]{\@secondoftwo}%
\providecommand \translation [1]{[#1]}%
\providecommand \BibitemOpen [0]{}%
\providecommand \bibitemStop [0]{}%
\providecommand \bibitemNoStop [0]{.\EOS\space}%
\providecommand \EOS [0]{\spacefactor3000\relax}%
\providecommand \BibitemShut  [1]{\csname bibitem#1\endcsname}%
\let\auto@bib@innerbib\@empty
\bibitem [{\citenamefont {Edelsbrunner}\ and\ \citenamefont {Harer}(2010)}%
  ]{EdelsbrunnerBook2010}%
  \BibitemOpen
  \bibfield  {author} {\bibinfo {author} {\bibfnamefont {H.}~\bibnamefont
  {Edelsbrunner}}\ and\ \bibinfo {author} {\bibfnamefont {J.~L.}~\bibnamefont
  {Harer}},\ }
  \href {https://bookstore.ams.org/mbk-69} {\bibfield  {title} {\bibinfo
  {title} {Computational topology: an introduction }}}%
  \bibfield  {publisher} {\bibinfo {publisher} {(American Mathematical
  Society}},\ \bibinfo {year} {2010)}
  \BibitemShut {NoStop}%

  \end{thebibliography}

\clearpage
\newpage

\onecolumngrid

\begin{center}
  \textbf{\large \hspace{5pt} SUPPLEMENTAL MATERIAL \\ \vspace{0.2cm}  Grain-Growth Stagnation from Vacancy-Diffusion-Limited Disconnection Climb \\}
\vspace{0.4cm}   Maik Punke,$^{1}$, Abel H. G. Milor,$^{1}$, Marco Salvalaglio$^{1,2,*}$ \\[.1cm]
  {\itshape \small
  ${}^1$Institute  of Scientific Computing,  TU  Dresden,  01062  Dresden,  Germany
  \\ 
  ${}^2$Dresden Center for Computational Materials Science, TU Dresden, 01062 Dresden, Germany
 }
\vspace{0.5cm}
\end{center}

\setcounter{equation}{0}
\setcounter{figure}{0}
\setcounter{table}{0}
\setcounter{page}{1}

\renewcommand{\thesection}{S-\Roman{section}}
\renewcommand{\theequation}{S-\arabic{equation}}
\renewcommand{\thefigure}{S-\arabic{figure}}
\renewcommand{\bibnumfmt}[1]{[S#1]}
\renewcommand{\citenumfont}[1]{S#1}

\section{S-I.\ \ Local Averaging}\label{section:averaging}
The coarse-graining operator $\langle \cdot \rangle$, employed to compute the local average density $\langle \psi \rangle$ and in the dynamics, denotes a weighted spatial averaging of the argument over the extent of a crystalline unit cell associated with the underlying square lattice symmetry. This averaging is implemented via convolution with a Gaussian kernel whose standard deviation is chosen to match the characteristic length of a unit cell $p=2\pi$ and reads $\langle \psi \rangle = \mathfrak{F}^{-1}\left[\mathrm{e}^{-\mathbf{k}^2 p^2/2} \mathfrak{F}[\psi]\right]$ with spatial Fourier vector $\mathbf{k}$ and (inverse) spatial Fourier transform $\mathfrak{F}^{(-1)}[f]$ of the periodic function $f$.
\section{S-II.\ \ Numerical Methods}\label{section:num}
The system of equations~(3) in the main text is solved numerically using a Fourier-based pseudo-spectral method enforcing periodic boundary conditions. Time integration is performed using an implicit–explicit (IMEX) approach:  the velocity field $\mathbf{v}$ is updated using a fully implicit first-order method, while the evolution of the order parameter $\psi$ is advanced with a first-order IMEX scheme (the additional $\nabla^2 \langle \psi \rangle$ term can be treated time-implicitly, further enhancing numerical stability). Numerical convergence is ensured by employing sufficiently fine spatial and temporal discretizations, with grid spacings $\Delta x=\Delta y=p/7$ and a time step size of $\Delta t=1/2$. The implementation is accelerated on graphics processing units (GPUs), with forward and inverse Fourier transforms computed using the cuFFT library.

\section{S-III.\ \ Grain Counting}\label{section:counting}
\begin{figure}[t]
\centering
\includegraphics[width=\linewidth]{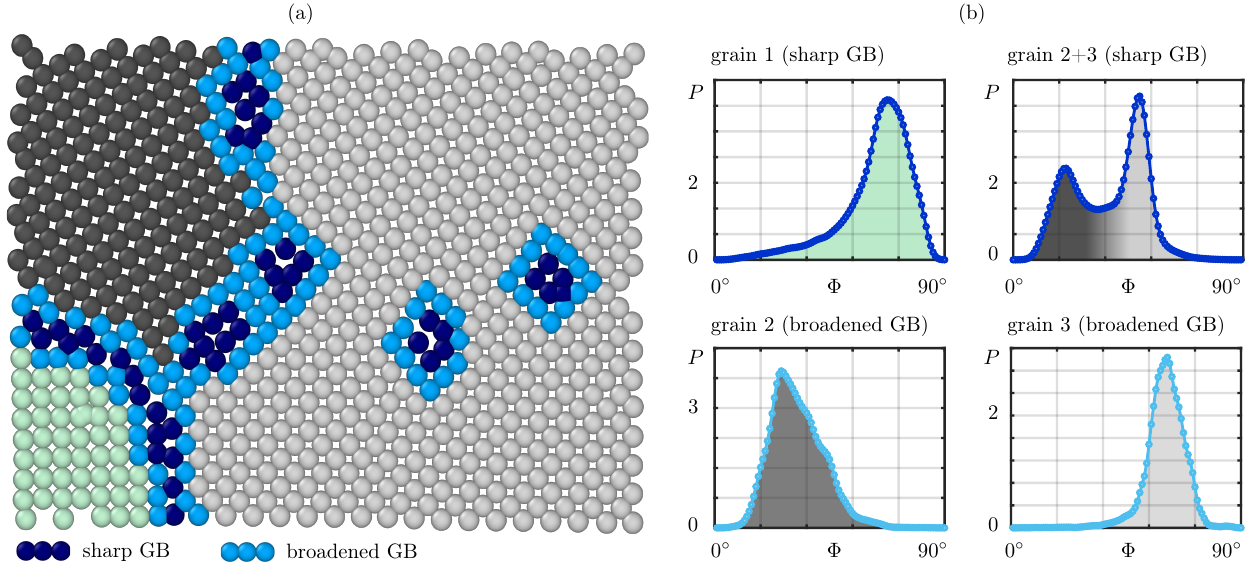}
\caption{\textit{Grain counting routine based on the local orientations of individual grains.} (a) Close-up of a polycrystalline configuration containing three individual grains (colored green, dark gray, and light gray, respectively). Grain boundaries and defects identified by applying a sharp threshold to the locally averaged density field $\langle \psi \rangle$ are highlighted in dark blue; the light blue regions indicate the broadened grain-boundary structure obtained by morphological opening.
(b) Probability distributions $P(\Phi)$ (in $\%$) of the local orientation field $\Phi$ evaluated within each connected component, shown for the sharp grain-boundary detection (top row) and for the broadened grain-boundary structure (bottom row).} 
\label{fig:SMgrainCounting}
\end{figure}

To evaluate the mean-squared grain size during polycrystalline coarsening, we employ two independent grain-counting methods.  
In Fig.~\ref{fig:SMgrainCounting}, we illustrate an approach based on the local orientations of individual grains. First, grain boundaries and isolated defects are identified by applying a threshold to the locally averaged density field $\langle \psi \rangle$, which are shown as dark blue atoms in Fig.~\ref{fig:SMgrainCounting}a. This procedure defines connected components: the first connected component corresponds to grain~1 (shown in green), while the second connected component contains grains~2 and~3 (shown in dark and light gray, respectively).
Next, we compute the local orientation field $\Phi$ and evaluate its probability distribution $P=P(\Phi)$ within each connected component, as shown in Fig.~\ref{fig:SMgrainCounting}b. If a component exhibits a single dominant orientation, it is identified as an individual grain, as in the case of grain~1. Otherwise, as for the combined grain~2+3 component, the boundary structure is progressively broadened using a morphological opening (indicated by light blue atoms in Fig.~\ref{fig:SMgrainCounting}a) until the connected component splits into regions of uniform orientation.
This iterative broadening procedure is repeated until all connected components contain only one dominant local orientation. In this way, low-angle grain boundaries—appearing as separated defect lines, such as those between grains~2 and~3—are distinguished from isolated defects within grains, for example, those inside grain~3.

Second, we apply persistent homology~\cite{EdelsbrunnerBook2010} to determine the number of grains. Similar to the method~\ref{fig:SMgrainCounting}, grain boundaries and isolated defects are identified by setting a threshold to the locally averaged density field $\langle \psi \rangle$ and using a smooth morphological opening.
Persistent homology captures the topological evolution over a parameter of the progressively broadened grain boundary structure (filtration), including the emergence of one-dimensional loops that reveal the grains.



For a small benchmark problem (approximately ten times smaller than the polycrystals used for the coarsening results in Fig.~2 in the main text), the average of the two methods yields a relative error of approximately $2.9\%$ in the number of grains compared with manual grain counting.

\end{document}